\documentclass[aps,prc,amsmath,preprint,superscriptaddress,showpacs]{revtex4}

\usepackage{bm,graphicx}
\begin{document}


\title{Complete set of polarization transfer coefficients
for the ${}^{3}{\rm He}(p,n)$ reaction at 346 MeV 
and 0 degrees}


\author{T.~Wakasa}
\email[]{wakasa@phys.kyushu-u.ac.jp}
\homepage[]{http://www.kutl.kyushu-u.ac.jp/~wakasa}
\affiliation{Department of Physics, Kyushu University,
Higashi, Fukuoka 812-8581, Japan}
\author{E.~Ihara}
\affiliation{Department of Physics, Kyushu University,
Higashi, Fukuoka 812-8581, Japan}
\author{M.~Dozono}
\affiliation{Department of Physics, Kyushu University,
Higashi, Fukuoka 812-8581, Japan}
\author{K.~Hatanaka}
\affiliation{Research Center for Nuclear Physics, Osaka University,
Ibaraki, Osaka 567-0047, Japan}
\author{T.~Imamura}
\affiliation{Department of Physics, Kyushu University,
Higashi, Fukuoka 812-8581, Japan}
\author{M.~Kato}
\affiliation{Research Center for Nuclear Physics, Osaka University,
Ibaraki, Osaka 567-0047, Japan}
\author{S.~Kuroita}
\affiliation{Department of Physics, Kyushu University,
Higashi, Fukuoka 812-8581, Japan}
\author{H.~Matsubara}
\affiliation{Research Center for Nuclear Physics, Osaka University,
Ibaraki, Osaka 567-0047, Japan}
\author{T.~Noro}
\affiliation{Department of Physics, Kyushu University,
Higashi, Fukuoka 812-8581, Japan}
\author{H.~Okamura}
\affiliation{Research Center for Nuclear Physics, Osaka University,
Ibaraki, Osaka 567-0047, Japan}
\author{K.~Sagara}
\affiliation{Department of Physics, Kyushu University,
Higashi, Fukuoka 812-8581, Japan}
\author{Y.~Sakemi}
\affiliation{Cyclotron and Radioisotope Center, Tohoku University,
Sendai, Miyagi 980-8578, Japan}
\author{K.~Sekiguchi}
\affiliation{RIKEN Nishina Center,
Wako, Saitama 351-0198, Japan}
\author{K.~Suda}
\affiliation{Research Center for Nuclear Physics, Osaka University,
Ibaraki, Osaka 567-0047, Japan}
\author{T.~Sueta}
\affiliation{Department of Physics, Kyushu University,
Higashi, Fukuoka 812-8581, Japan}
\author{Y.~Tameshige}
\affiliation{Research Center for Nuclear Physics, Osaka University,
Ibaraki, Osaka 567-0047, Japan}
\author{A.~Tamii}
\affiliation{Research Center for Nuclear Physics, Osaka University,
Ibaraki, Osaka 567-0047, Japan}
\author{H.~Tanabe}
\affiliation{Department of Physics, Kyushu University,
Higashi, Fukuoka 812-8581, Japan}
\author{Y.~Yamada}
\affiliation{Department of Physics, Kyushu University,
Higashi, Fukuoka 812-8581, Japan}

\date{\today}

\begin{abstract}
 We report measurements of the cross-section and a complete
set of polarization transfer coefficients for the
${}^{3}{\rm He}(p,n)$ reaction at a bombarding energy
$T_p$ = 346 MeV and a reaction angle $\theta_{\rm lab}$ =  $0^{\circ}$.
 The data are compared with the corresponding free nucleon-nucleon 
values on the basis of the predominance of quasi-elastic scattering 
processes.
 Significant discrepancies have been observed in the 
polarization transfer $D_{LL}(0^{\circ})$, which are presumably the result of the three-proton $T$ = $3/2$ resonance.
 The spin--parity of the resonance is estimated to be $1/2^-$, 
and the distribution is consistent with previous results 
obtained for the same reaction at $T_p$ = 48.8 MeV.
\end{abstract}


\maketitle

\section{INTRODUCTION}
\label{sec:intro}

 The search for resonances in the three-nucleon system 
with isospin $T$ = $3/2$ has a long and interesting story.
 Evidence for $T$ = $3/2$ (three-proton) resonance was 
found in the ${}^{3}{\rm He}(p,n)$ reaction at 
the proton incident energy $T_p$ = 48.8 MeV \cite{prl_23_1181_1969}.
 The maximum enhancement from the four-body phase space 
was observed at the excitation energy $E_x$ = $9\pm 1$ MeV 
with a width $\Gamma$ = $10.5\pm 1$ MeV in the three-proton system
(where $E_x$ = 0 at the three-proton rest mass energy).
 Similar enhancements have been observed for
measurements at $T_p$ = 24.9 MeV \cite{plb_29_573_1969} and 197 MeV
\cite{prc_58_645_1998}.
 However, these studies have attributed the observed 
enhancements to a 
${}^{1}{\rm S}_0$ two-proton final state interaction (FSI), 
rather than a three-proton resonance.
 In the impulse approximation, the $(p,n)$ reaction occurs 
only on the unpaired neutron in ${}^{3}{\rm He}$ and the two 
target protons coupled to the ${}^{1}{\rm S}_0$ state act 
as spectator ${}^{2}{\rm He}$.
 Thus, the final state consists of three particles of 
$p$, $n$, and ${}^{2}{\rm He}$, and the three-body phase space 
calculations reasonably explain the experimental 
enhancements from the four-body phase space.
 Therefore, the interpretation of the enhancements in the 
cross-section remains controversial.

 Recently three-neutron $T$ = $3/2$ resonances have been studied in 
the framework of configuration-space Faddeev equations 
\cite{prc_60_024002_1999,prc_66_054001_2002}.
 Unfortunately, these studies have indicated the 
absence of three-neutron resonances.
 However, the calculations did not include three-nucleon 
force effects.
 Thus, if there exists a three-nucleon resonance with $T$ = $3/2$, 
the properties of this resonance would yield valuable information 
on the $T$ = $3/2$ three-nucleon forces \cite{prc_64_014001_2001}, 
on which experimental data are scarce.

 In this article, we present the double-differential cross-section and a complete set of polarization transfer coefficients 
for the ${}^{3}{\rm He}(p,n)$ reaction at $T_p$ = 346 MeV and 
a reaction angle $\theta_{\rm lab}$ = $0^{\circ}$.
 Polarization transfer coefficients are sensitive to the 
spin--parity $J^{\pi}$ of an excited state \cite{prc_26_727_1982}, 
and thus they 
are sensitive to the presence of a resonance that has a 
fixed $J^{\pi}$, as was demonstrated for the spin-dipole 
resonances in ${}^{12}{\rm N}$
\cite{jpsj_77_014201_2008}.
 It should be noted that the polarization transfer data
for the ${}^{2}{\rm H}(p,n)$ reaction at the same incident energy 
and $\theta_{\rm lab}$ = $0^{\circ}$--$27^{\circ}$
have been well described by the free nucleon--nucleon ({\it NN}\,) 
values in the optimum frame 
\cite{npa_631_757c_1998,prc_59_3177_1999,prc_69_044602_2004}.
 Thus, comparison of the measured ${}^{3}{\rm He}(p,n)$ data 
with the corresponding free {\it NN} values enables us to assess 
whether there exists a three-nucleon resonance.
 Significant differences have been observed in our data which 
can be interpreted as evidence for three-proton 
resonance effects.
 The resonance properties 
have been discussed by comparing with 
distorted wave impulse approximation (DWIA) calculations.

\section{EXPERIMENTAL METHODS}
\label{sec:exp}

 The data were obtained with a 
neutron time-of-flight (NTOF) system \cite{nima_369_120_1996} 
and a neutron detector and polarimeter NPOL3 \cite{nima_547_569_2005} 
at the Research Center for Nuclear Physics (RCNP), Osaka University.
 The experimental setup and procedure were 
similar to those reported previously
\cite{jpsj_77_014201_2008,plb_645_402_2007,plb_656_38_2007}.
 Thus, in the following, we describe
the detector system only briefly and discuss experimental details relevant
to the present experiment.

\subsection{Polarized proton beam}

 A high-intensity polarized ion source (HIPIS) at RCNP 
\cite{nima_384_575_1997} was used to produce the polarized 
proton beam.
 The beam polarization direction was reversed every 5 s 
by selecting rf transitions in order to minimize 
geometrical false-asymmetries.
 The beam was accelerated up to $T_p$ = 346 MeV 
by using the AVF and Ring cyclotrons.
 One out of seven beam pulses was selected before injecting into 
the Ring cyclotron, which then yielded a beam pulse period of 431 ns.
 This pulse selection reduces the wraparound of slow 
neutrons from preceding beam pulses.
 The single-turn extraction was maintained during the measurement 
in order to keep the beam polarization.

 The superconducting solenoid magnets SOL1 and SOL2
located in the injection line from the AVF to Ring 
cyclotrons were used to precess the proton spin direction.
 Each magnet can rotate the direction of the polarization 
vector from the normal $\hat{N}$ into sideways $\hat{S}$ 
directions.
 These two magnets were separated by a bending angle 
of $45^{\circ}$, and the spin precession angle in this 
bending magnet was about $85.8^{\circ}$.
 Thus, the longitudinal ($\hat{L}$) and sideways ($\hat{S}$)
polarized proton beams can be obtained at the exit 
of SOL2 by using the SOL1 and SOL2 magnets, respectively.

 The beam polarization was continuously monitored
by two sets of beam-line polarimeters, BLP1 and BLP2.
 These two polarimeters were separated by a 
bending angle of $98^{\circ}$, allowing simultaneous 
determination of all of the components of the polarization 
vector including the longitudinal component.
 Each polarimeter consists of four pairs of conjugate-angle plastic 
scintillators.
 The $\vec{p}+p$ elastic scattering was used as the analyzing 
reaction, and a self-supporting ${\rm CH_2}$ target with a 
thickness of 1.1 ${\rm mg/cm^2}$ was used as the hydrogen target.
 The elastically scattered and recoiled protons were detected 
in kinematical coincidence with a pair of scintillators.
 The typical magnitude of the beam polarization was about 0.60.

\subsection{${}^{3}{\rm He}$ target}

 The ${}^{3}{\rm He}$ target was prepared as a high-pressure 
cooled gas target by using a target system developed for a 
liquid ${\rm H}_2$ target \cite{mpla_18_322_2003}.
 This target was operated at temperatures down to 25 K and 
at absolute pressures up to 2.5 atm.
 Both the cell temperature and pressure were continuously 
monitored during the experiment, and the typical areal 
density was about 11 ${\rm mg/cm^2}$.
 The gas cell windows were made of 12-$\mu {\rm m}$-thick 
Alamid foil.
 Background (empty-target) spectra were also measured 
in order to subtract the contribution from the Alamid 
windows.
 We also measured data with ${\rm D}_2$ in the target 
cell at 25 and 50 K under 2.5 atm, 
and compared these with the data for a 
solid ${\rm CD_2}$ target having a thickness of 228 ${\rm mg/cm^2}$.
 After correcting for the difference in areal 
density of the gaseous and solid targets, these data 
agreed within the systematic uncertainty 
associated with the areal density of the gaseous target 
($\simeq$ 7\%), suggesting that the performance of the 
gaseous target is well understood.

\subsection{Neutron spin rotation magnet and NPOL3}

 A dipole magnet (NSR magnet) located at the entrance 
of the time-of-flight (TOF) 
tunnel was used to precess the neutron polarization 
vector from the longitudinal $\hat{L}'$ into normal $\hat{N}'$ 
directions so as to make the longitudinal component measurable 
with NPOL3.
 In the measurement of the longitudinal component of the 
neutron polarization, the NSR magnet was excited so that 
the precession angle for the neutron corresponding to the energy
transfer $\omega_{\rm lab}$ = 5 MeV became $90^{\circ}$.
 Corrections for the over- and under-precessions to the 
lower and higher energy neutrons were made to account 
for the mixing between the longitudinal and normal components.

 Neutrons were measured by the NPOL3 system \cite{nima_547_569_2005} 
with a 70~m flight path length.
 The NPOL3 system consists of 
three planes of neutron detectors.
 The first two planes (HD1 and HD2), which act as neutron detectors 
and neutron polarization analyzers, are made of 20 
sets of one-dimensional position-sensitive plastic scintillators (BC408) 
with a size of 100 $\times$ 10 $\times$ 5 ${\rm cm}^3$.
 The last plane (NC), which serves as a catcher for the 
particles scattered by HD1 or HD2, is 
made of a two-dimensional position-sensitive 
liquid scintillator (BC519) 
with a size of 100 $\times$ 100 $\times$ 10 ${\rm cm}^3$.
 Each of the three neutron detectors has an effective detection 
area of 1 ${\rm m^2}$.

\section{DATA REDUCTION}
\label{sec:reduction}

\subsection{Neutron detection efficiency}

 The neutron detection efficiency of NPOL3 
(HD1 and HD2) was determined 
using the ${}^{7}{\rm Li}(p,n){}^{7}{\rm Be}({\rm g.s.}+0.43\,{\rm
MeV})$ reaction at $\theta_{\rm lab}$ = $0^{\circ}$ 
whose cross-section is known at 
$T_p$ = 80--795 MeV \cite{prc_41_2548_1990}.
 The result is 0.053$\pm$0.003 
where the uncertainty comes mainly from the 
uncertainty in both the ${}^{7}{\rm Li}$ cross-section 
(3\%) and the thickness of the ${}^{7}{\rm Li}$ target (3\%).

\subsection{Effective analyzing power}

 The neutron polarization was analyzed by using the $\vec{n}+p$
scattering in either neutron detector HD1 or HD2, and the recoiled 
protons were detected with neutron detector NC.
 The effective analyzing power $A_{y;{\rm eff}}$ of NPOL3 
was determined by using polarized neutrons from the 
Gamow--Teller (GT) ${}^{2}{\rm H}(p,n)pp({}^{1}{\rm S}_0)$ reaction at 
$T_p$ = 346 MeV and $\theta_{\rm lab}$ = $0^{\circ}$.
 We used two kinds of polarized protons 
with normal ($p_N$) and longitudinal ($p_L$)
polarizations.
 The corresponding neutron polarizations at $0^{\circ}$ 
become $p_N'=p_ND_{NN}(0^{\circ})$ and 
$p_L'=p_L D_{LL}(0^{\circ})$.
 The resulting asymmetries measured by NPOL3 
are
\begin{subequations}
\label{eq_asym}
\begin{eqnarray}
\epsilon_N & = & 
   p_N'A_{y;{\rm eff}}=p_ND_{NN}(0^{\circ})A_{y;{\rm eff}}\ ,\\
\epsilon_L & = & 
   p_L'A_{y;{\rm eff}}=p_LD_{LL}(0^{\circ})A_{y;{\rm eff}}\ .
\label{eq_asym_dll}
\end{eqnarray}
\end{subequations}
 Because the polarization transfer coefficients for the GT
transition satisfy \cite{jpsj_73_1611_2004}
\begin{equation}
2D_{NN}(0^{\circ})+D_{LL}(0^{\circ})=-1\ ,
\label{eq:gt_dii}
\end{equation}
$A_{y;{\rm eff}}$ can be expressed by using 
Eqs.~(\ref{eq_asym}) and (\ref{eq:gt_dii}) as 
\begin{equation}
A_{y;{\rm eff}} = -\left(
2\frac{\epsilon_N}{p_N}+\frac{\epsilon_L}{p_L}
\right)\ .
\end{equation}
 Therefore, the $A_{y;{\rm eff}}$ value can be obtained without
knowing a priori the values of $D_{ii}(0^{\circ})$, and the 
result is $A_{y;{\rm eff}}$ = 0.130 $\pm$ 0.004
where the uncertainty is statistical.

\begin{figure}
\begin{center}
\includegraphics[width=0.7\linewidth,clip]{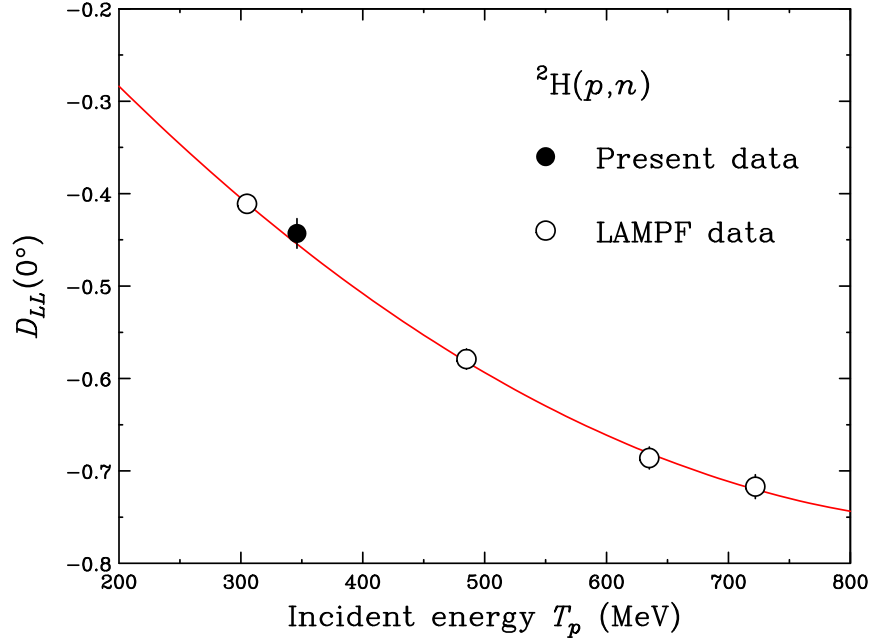}
\end{center}
\caption{(Color online)
 Polarization transfer $D_{LL}(0^{\circ})$ 
for the ${}^{2}{\rm H}(p,n)$ reaction 
at $0^{\circ}$ as a function of incident 
energy $T_p$.
 The filled circle is the present result, while 
the open circles show the data by 
McNaughton {\it et al.} \protect{\cite{prc_45_2564_1992}}. 
 The solid curve represents the fit with a second order 
polynomial.}
\label{fig1}
\end{figure}

 The $D_{LL}(0^{\circ})$ values of the 
${}^{2}{\rm H}(p,n)pp$ reaction at $T_p$ = 305--788 MeV 
have been reported by McNaughton {\it et al.} \cite{prc_45_2564_1992};
the results are shown in Fig.~\ref{fig1} with
open circles.
 The error bars represent both statistical and systematic
uncertainties.
 The solid curve is the result of fitting with a second order 
polynomial.
 The $D_{LL}(0^{\circ})$ value at $T_p$ = 346 MeV, which is
determined from Eq.~(\ref{eq_asym_dll}) by using the 
$A_{y;{\rm eff}}$ value obtained, is indicated in
Fig.~\ref{fig1} by the filled circle.
 Our $D_{LL}(0^{\circ})$ value is consistent with the 
energy dependence predicted on the basis of previous data, 
demonstrating the reliability of our calibrations.

\subsection{Background subtraction}
 
 Observables for the ${}^{3}{\rm He}(p,n)$ reaction were 
extracted through a cross-section-weighted subtraction 
of the observables for the empty target from the observables 
for the full target as 
\begin{subequations}
\label{eq_sub}
\begin{eqnarray}
\sigma_{{}^{3}{\rm He}} & = & 
   \sigma_{\rm Full}-\sigma_{\rm Empty}\ ,\\
D_{{}^{3}{\rm He}} & = & 
\frac{D_{\rm Full}-fD_{\rm Empty}}{1-f}\ ,
\end{eqnarray}
\end{subequations}
where $\sigma$ represents the cross-section, 
$D$ is one of the polarization transfer coefficients 
$D_{ii}(0^{\circ})$, 
and $f=\sigma_{\rm Empty}/\sigma_{\rm Full}$.
 The fraction $f$ was estimated by using the cross-sections
based on the nominal target thicknesses and integrated 
beam current.

\begin{figure}
\begin{center}
\includegraphics[width=0.7\linewidth,clip]{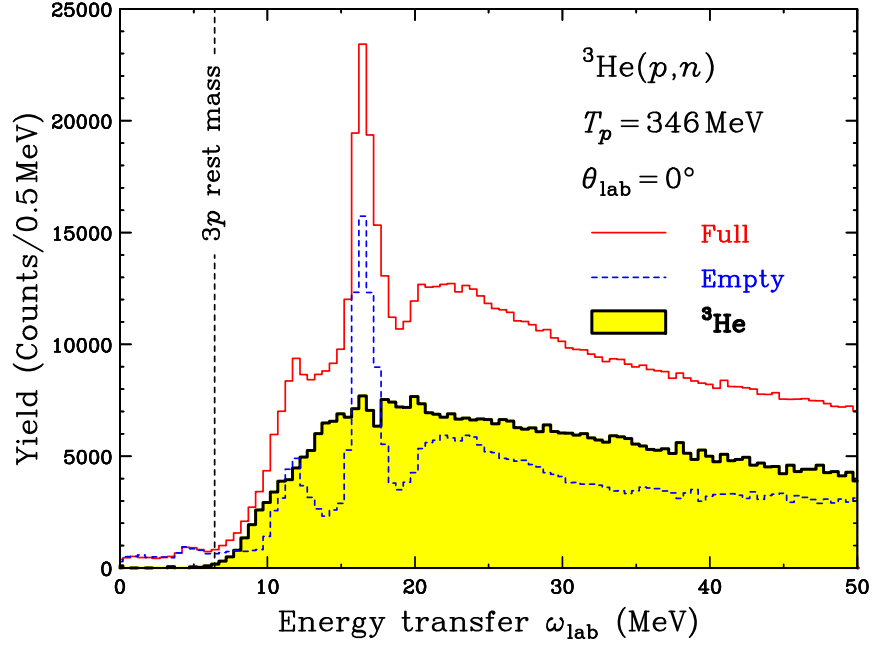}
\end{center}
\caption{(Color online)
 Energy transfer spectra with an empty target (dashed histogram) 
and a target filled with ${}^{3}{\rm He}$ gas (thin-solid histogram) 
for the $(p,n)$ reaction 
at $T_p$ = 346 MeV and $\theta_{\rm lab}$ = $0^{\circ}$.
 The narrow peaks are from $(p,n)$ reactions on the elements 
of the Alamid windows.
 The filled thick-solid histogram shows the spectra 
for the ${}^{3}{\rm He}(p,n)$ reaction obtained by the 
subtraction of Eq.~(\ref{eq_sub}).}
\label{fig2}
\end{figure}

 Figure~\ref{fig2} shows a representative set of 
spectra as a function of $\omega_{\rm lab}$.
 In both the full and empty target spectra, 
narrow peaks are observed at $\omega_{\rm lab}$ = 12 and 17 
MeV and a broad bump is centered near 22 MeV.
 The narrow peaks result from the 
${}^{14}{\rm N}(p,n){}^{14}{\rm O}(2^+,7.7\,{\rm MeV})$ and 
${}^{12}{\rm C}(p,n){}^{12}{\rm N}(1^+,{\rm g.s.})$ 
reactions on the Alamid windows.
 The broad bump is mainly due to the spin-dipole resonances 
in ${}^{12}{\rm N}$ excited by the $(p,n)$ reaction on 
${}^{12}{\rm C}$ in Alamid.
 The signal-to-background ratio, integrated up to $\omega_{\rm lab}$ 
= 50 MeV, is about 1.3, which is significantly better than that 
obtained in the previous experiment (0.17) at $T_p$ = 197 MeV 
\cite{prc_58_645_1998}.
 This is mainly thanks to the use of the relatively thin
Alamid foil for the target windows.

 The filled histogram in Fig.~\ref{fig2} shows the 
subtraction results. 
 The background contributions including narrow peaks and the 
broad bump are successfully subtracted without adjusting the 
relative normalization, 
demonstrating the reliability of our measurements.
 The vertical dashed line represents the energy transfer 
for the three-proton rest system.
 Because there is no bound state in the three-proton system, 
the spectrum of the ${}^{3}{\rm He}(p,n)$ reaction shows a 
rise due to this energy transfer.

\section{RESULTS}
\label{sec:results}

\begin{figure}
\begin{center}
\includegraphics[width=0.7\linewidth,clip]{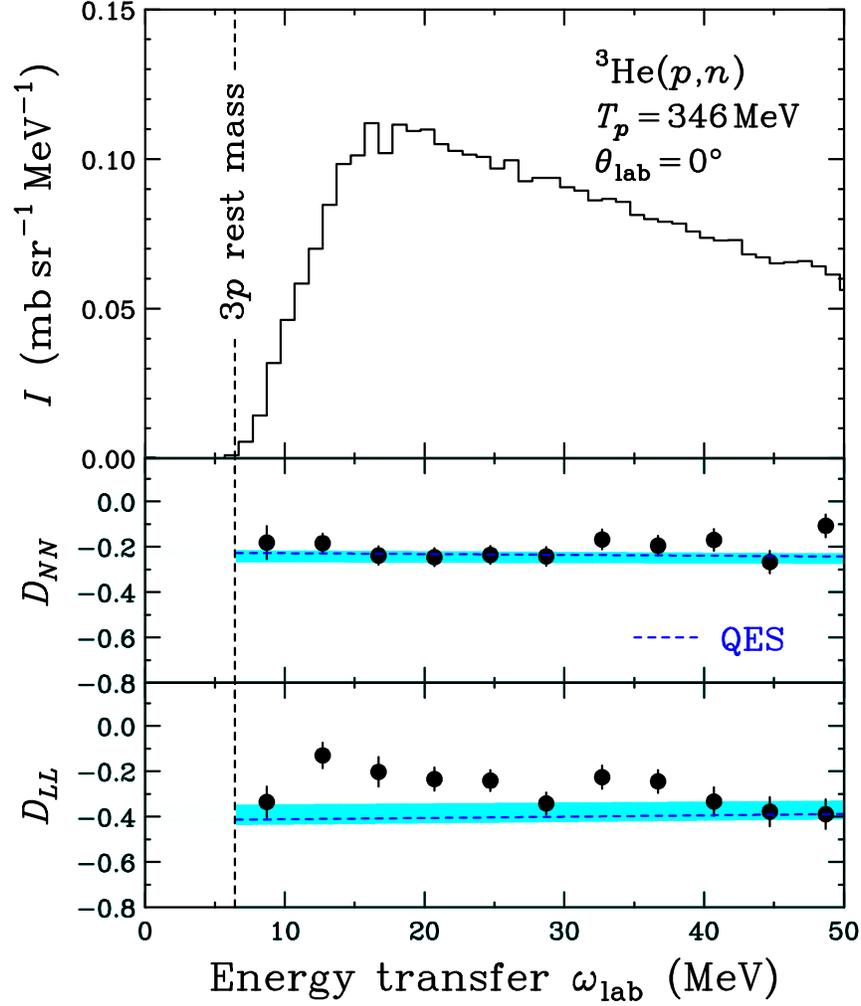}
\end{center}
\caption{(Color online)
 The double-differential cross-section $I$ (top panel) and complete set of polarization transfer coefficients, 
$D_{NN}(0^{\circ})$ (middle panel) and $D_{LL}(0^{\circ})$ 
(bottom panel), for the ${}^{3}{\rm He}(p,n)$ reaction
at $T_p$ = 346 MeV and $\theta_{\rm lab}$ = $0^{\circ}$.
 The dashed curves are the corresponding free {\it NN} values
with the {\sc fa07} phase-shift solution 
\protect{\cite{prc_76_025209_2007}}.
 The shaded bands represent the uncertainty in the free 
{\it NN} values estimated by using the up-to-date 
{\it NN} potentials, as described in the text.}
\label{fig3}
\end{figure}

 Figure~\ref{fig3} shows the double-differential 
cross-section $I$ and the complete set of polarization 
transfer coefficients $D_{NN}(0^{\circ})$ and $D_{LL}(0^{\circ})$
for the ${}^{3}{\rm He}(p,n)$ reaction 
at $T_p$ = 346 MeV and $\theta_{\rm lab}$ = $0^{\circ}$.
 The data for the cross-section are binned in 1 MeV intervals, 
while the data for $D_{ii}(0^{\circ})$ are binned in 
2 MeV intervals to reduce statistical fluctuations.
 The slope of the cross-section spectrum near the threshold 
is primarily determined by the phase space factor, and 
is consistent with that at $T_p$ = 197 MeV 
\cite{prc_58_645_1998}.
 Palarczyk {\it et al.} \cite{prc_58_645_1998} reported phase-space 
calculations and showed that the 
increase in cross-section near the threshold is reproduced 
better by three-body than four-body phase space calculations.
 This means that the present ${}^{3}{\rm He}(p,n)$ reaction 
can be described by $n(p,n)p$ quasi-elastic scattering 
on the neutron in ${}^{3}{\rm He}$, whereas the two protons 
in the relative ${}^{1}{\rm S}_0$ state act as a spectator.
 Thus, the measured polarization transfer coefficients 
are expected to be well described by the corresponding
free {\it NN} values.
 
 The dashed curves in Fig.~\ref{fig3} represent the 
corresponding free {\it NN} values with the {\sc fa07} phase-shift 
solution \cite{prc_76_025209_2007} of the on-line 
Scattering Analysis Interactive Dial-in (SAID) Facility \cite{said}.
 The uncertainty in the free {\it NN} values was evaluated 
by using the modern nucleon-nucleon {\it NN} potentials of 
AV18 \cite{prc_51_38_1995}, 
CD Bonn \cite{prc_63_024001_2001}, 
Nijmegen 93 \cite{prc_49_2950_1994}, and 
Paris \cite{prc_21_861_1980}.
 The results are represented in Fig.~\ref{fig3} by
the shaded bands.
 The measured $D_{NN}(0^{\circ})$ values are close to the
corresponding free {\it NN} values.
 This supports the predominance of quasi-elastic scattering 
processes in this reaction.
 However, significant discrepancies 
are observed in $D_{LL}(0^{\circ})$.
 The discrepancies at $\omega_{\rm lab}$ $\gtrsim$ 30 MeV 
might be due to the energy dependence of $A_{y;{\rm eff}}$ 
which was neglected in the present analysis.

 The present analysis was based 
on the assumption of simple 
quasi-elastic scattering, and thus the discrepancies at 
$\omega_{\rm lab}$ $\lesssim$ 30 MeV
do not necessarily evidence
$T$ = $3/2$ three-proton resonance.
 One possibility is the effects of the $D$-state 
in ${}^{3}{\rm He}$.
 The deuteron $D$-state effects for ${}^{2}{\rm H}(p,n)$ at 
the same incident energy were studied by Sakai {\it et al.} 
\cite{npa_631_757c_1998} 
using the plane wave impulse approximation code developed 
by Itabashi {\it et al.}
\cite{ptp_91_69_1994}.
 They concluded that the effects to the cross section at 
$\theta_{\rm lab}$ =  $0^{\circ}$ are negligible at small 
energy transfers where the ${}^{1}{\rm S}_0$ FSI process 
is dominant.
 We performed the calculations using the same code in order 
to investigate the $D$-state effects to the 
polarization transfer coefficients at energy transfers 
up to 50 MeV.
 The $D$-state contributions to the cross section 
become appreciable beyond the FSI region 
as increasing the energy transfer.
 However, their effects to the polarization transfer 
coefficients are very small, namely, 
less than 0.04 at $\omega_{\rm lab}$ $<$ 50 MeV.
 Therefore, we expect that the $D$-state effects 
to $D_{NN}(0^{\circ})$ and $D_{LL}(0^{\circ})$ 
for ${}^{3}{\rm He}(p,n)$ 
are also small, and thus 
it is interesting to investigate the 
discrepancies by assuming the effects due to the 
$T$ = $3/2$ three-proton resonance contribution.
 In the following, we performed DWIA 
calculations in order to
determine the $J^{\pi}$ and strength of the resonance.

\section{DISCUSSION}
\label{sec:discuss}

\subsection{DWIA calculations}

 We performed full microscopic DWIA calculations by 
using the computer code {\sc dw81} \cite{dw81},
which treats the knock-on exchange amplitude exactly.
 The final states with $J^{\pi}$ = $1/2^{\pm}$ and $3/2^{\pm}$ 
were investigated \cite{prc_66_054001_2002}.
 The one-body density matrix elements for the 
transitions to these states by ${}^{3}{\rm He}(p,n)$ were
obtained with the shell-model (SM) code {\sc oxbash}
\cite{oxbash}.
 The SM calculations were performed in the $0s$-$0p$-$1s0d$-$0f1p$
configuration space by using the phenomenological effective 
interaction optimized for $A$=3 by Hosaka, Kubo, and Toki 
\cite{npa_444_76_1985}.
 For each transition, only the lowest-energy state was
investigated by DWIA calculations because the $D_{ii}(0^{\circ})$ 
values are primarily determined by $J^{\pi}$ \cite{prc_26_727_1982}.
 The single-particle radial wave functions were assumed to 
have a harmonic oscillator shape with the range parameter 
$b$ = 1.67 fm \cite{prc_39_336_1989}.
 The optical model potential (OMP) was deduced from the 
global OMPs optimized for ${}^{4}{\rm He}$ in the 
proton energy range $T_p$ = 156--1728 MeV \cite{prc_73_024608_2006}.
 The {\it NN} {\it t}-matrix parameterized by Franey and Love 
at 325 MeV \cite{prc_31_488_1985} was used.

 The DWIA calculation results are summarized in 
Table~\ref{table:dwia}.
 In order to reproduce the observed $D_{ii}(0^{\circ})$ values,
the $D_{NN}(0^{\circ})$ value of the resonance should be
close to the observed value of $\sim -0.2$, whereas the
$D_{LL}(0^{\circ})$ value should be significantly larger 
than the observed value of $\sim -0.3$.
 This constraint is satisfied only in the case of 
$J^{\pi}$ = $1/2^-$.
 Thus, in the following, we deduce the strength distribution 
of the resonance by assuming $J^{\pi}$ = $1/2^-$.

\subsection{Resonance contributions}

 Here we assume that the observed ${}^{3}{\rm He}(p,n)$ 
cross-section comprises an incoherent sum of contributions 
from quasi-elastic scattering and the resonance
with $J^{\pi}$ = $1/2^-$.
 Thus, the observed $D_{ii}(0^\circ)$ values can be 
expressed by using the observed cross-section 
$\sigma(0^{\circ})$ as 
\begin{subequations}
\label{eq:dii_decom}
\begin{eqnarray}
D_{NN}(0^{\circ}) & = & 
\frac{\sigma^{1/2^-}(0^{\circ})D_{NN}^{1/2^-}(0^{\circ})+
(\sigma(0^{\circ})-\sigma^{1/2^-}(0^{\circ}))
D_{NN}^{\rm QES}(0^{\circ})}
{\sigma(0^{\circ})}\ ,\\
D_{LL}(0^{\circ}) & = & 
\frac{\sigma^{1/2^-}(0^{\circ})D_{LL}^{1/2^-}(0^{\circ})+
(\sigma(0^{\circ})-\sigma^{1/2^-}(0^{\circ}))
D_{LL}^{\rm QES}(0^{\circ})}
{\sigma(0^{\circ})}\ ,
\end{eqnarray}
\end{subequations}
where $D_{ii}^{1/2^-}(0^{\circ})$ and $D_{ii}^{\rm QES}(0^{\circ})$ 
are the $D_{ii}(0^{\circ})$ values for 
the resonance with $J^{\pi}$ = $1/2^-$ and 
the quasi-elastic scattering (free {\it NN} scattering), 
respectively, 
and $\sigma^{1/2^-}$ is the cross-section 
of the resonance.
 The shape of $\sigma^{1/2^-}$ is 
described by a Bright--Wigner (Lorentz) 
function, and threshold (phase-space) effects 
are taken into account.
 The center $\omega_0$, width $\Gamma$ (full width at half
maximum), and amplitude of $\sigma^{1/2^-}$ are determined 
to satisfy Eq.~(\ref{eq:dii_decom}) 
by using the $D_{ii}^{1/2^-}(0^{\circ})$ and $D_{ii}^{\rm QES}(0^{\circ})$ 
values evaluated from the DWIA calculations and 
free {\it NN} values, respectively.

\begin{figure}
\begin{center}
\includegraphics[width=0.7\linewidth,clip]{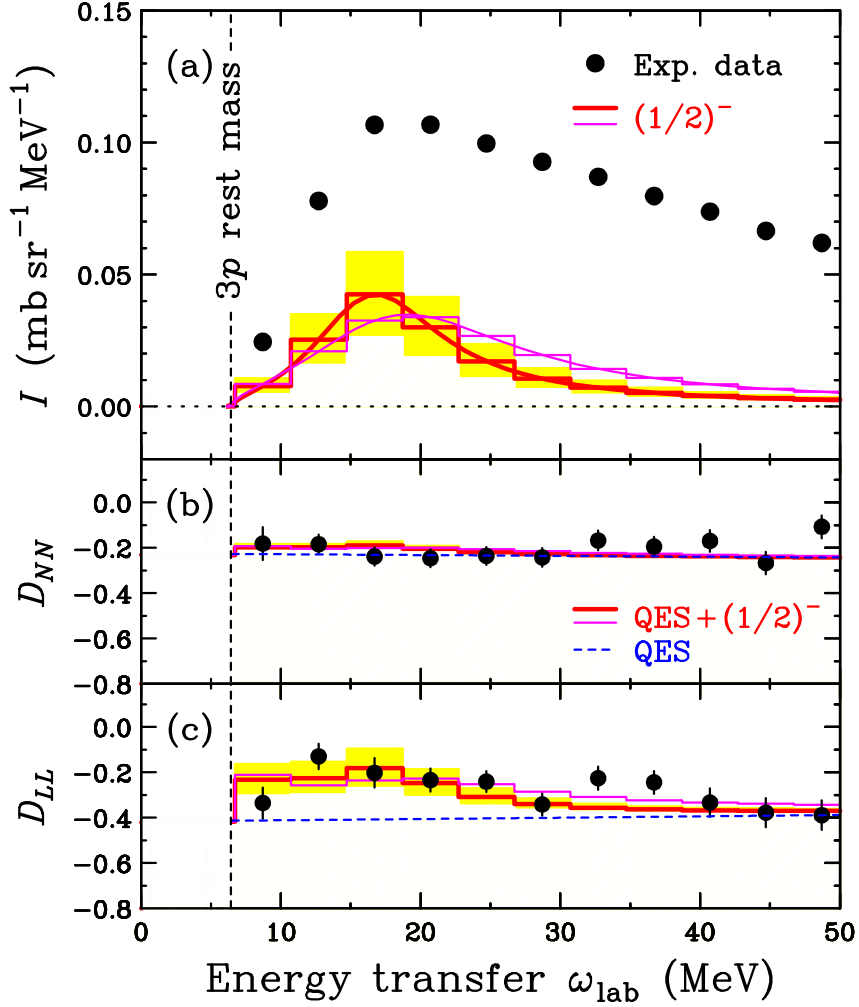}
\end{center}
\caption{(Color online)
(a) The estimated $J^{\pi}$ = $1/2^-$ $T$ = $3/2$ 
resonance cross-section (solid histograms and curves) compared 
with the total cross-section (filled circles) 
for the ${}^{3}{\rm He}(p,n)$ reaction 
at $T_p$ = 346 MeV and $\theta_{\rm lab}$ = $0^{\circ}$.
 The thick and thin lines are the results of fitting 
for $\omega_{\rm lab}$ $<$ 30 and 50 MeV, respectively.
(b) The $D_{NN}(0^{\circ})$ values including the 
$J^{\pi}$ = $1/2^-$ resonance contributions with 
Eq.~(\ref{eq:dii_decom}) (solid histogram) compared with the 
experimental data (filled circles).
 The band represents the uncertainty of the fitting 
for $\omega_{\rm lab}$ $<$ 30 MeV due to 
the uncertainty of the experimental data.
 The dashed curve represents the corresponding 
free {\it NN} values
with the {\sc fa07} phase-shift solution 
\protect{\cite{prc_76_025209_2007}}.
(c) Same as (b), but for $D_{LL}(0^{\circ})$.}
\label{fig4}
\end{figure}

 The thick and thin lines in Fig.~\ref{fig4} show the 
results of fitting for 
$\omega_{\rm lab}$ $<$ 30 and 50 MeV, respectively.
 The solid histograms and curves in the top panel 
represent the $\sigma^{1/2^-}$ values.
 The solid histograms in the lower two panels
are the $D_{ii}(0^{\circ})$ values evaluated by
Eq.~(\ref{eq:dii_decom}).
 The shaded bands represent the errors of the fitting results 
for $\omega_{\rm lab}$ $<$ 30 MeV
due to the uncertainty of the experimental data.
 The dashed curves show the $D_{ii}^{\rm QES}(0^{\circ})$ values 
of the quasi-elastic scattering contribution.
 By considering the contributions from $J^{\pi}$ = $1/2^-$, 
both the $D_{NN}(0^{\circ})$ and $D_{LL}(0^{\circ})$ values 
are well reproduced.

 The center of $\sigma^{1/2^-}$ is almost independent of the 
fitting region, and the results are $\omega_0$ = $16\pm 1$ 
and 
$17\pm 1$ MeV for the fitting regions 
of $\omega_{\rm lab}$ $<$ 30 and 50 MeV, respectively.
 However, the width depends on the fitting region and the 
results are $\Gamma$ = $11\pm 3$ and $19\pm 6$ MeV 
for $\omega_{\rm lab}$ $<$ 30 and 50 MeV, respectively.
 In the case of $\omega_{\rm lab}$ $<$ 50 MeV, the large
$\Gamma$ value of 19 MeV makes it difficult to interpret
this contribution as a resonance.
 On the contrary, if we adopt the results for 
$\omega_{\rm lab}$ $<$ 30 MeV where the systematic 
uncertainties of the data coming from the energy dependence 
of $A_{y;{\rm eff}}$ are small, 
the interpretation as a resonance is reasonable due to 
the relatively narrow width of 11 MeV.
 If we choose the excitation energy $E_x$ = 0 of the 
three-proton system to be at the three-proton rest mass energy,
$\omega_0$ = 16 MeV corresponds to $E_x$ = 10 MeV.
 Thus, it is very interesting to note that our results 
are consistent with the results of 
$E_x$ = $9\pm 1$ MeV and $\Gamma$ = $10.5\pm 1$ MeV 
for a possible $T=3/2$ resonance observed by the same reaction
\cite{prl_23_1181_1969}.
 Because the present analyses have been based on the 
simple quasi-elastic scattering mechanism, 
detailed theoretical investigations are highly required
in order to confirm the discrepancies observed in $D_{LL}(0^{\circ})$ 
as the three-proton resonance effects.

\section{SUMMARY AND CONCLUSION}
\label{summary}

 The cross-section and a complete set of polarization 
transfer coefficients were measured for the 
${}^{3}{\rm He}(p,n)$ reaction at $T_p$ = 346 MeV and 
$\theta_{\rm lab}$ = $0^{\circ}$.
 The data are compared with the corresponding free 
{\it NN} values under the assumption that
the quasi-elastic scattering processes are predominant.
 Significant deviations were observed from the 
corresponding {\it NN} values in the polarization 
transfer $D_{LL}(0^{\circ})$.
 These discrepancies can be attributed to the 
three-nucleon $T$ = $3/2$ resonance whose 
spin--parity $J^{\pi}$ is estimated to be $1/2^-$ in 
DWIA calculations.
 The center $\omega_0$ and the width $\Gamma$ 
of the resonance are estimated to be 
$\omega_0$ = $16\pm 1$ MeV 
and $\Gamma$ = $11\pm 3$ MeV by using the data
at $\omega_{\rm lab}$ $<$ 30 MeV whose systematic 
uncertainties are small.
 The estimated strength distribution 
is consistent with previous results obtained from
the same reaction at $T_p$ = 48.8 MeV.
 However, the present data are not conclusive evidence 
for the three-nucleon resonance, and call for the theoretical 
calculation which incorporates the Coulomb interaction and the
$T$ = $3/2$ three-nucleon forces in order to settle the interpretation
of the present data.
 
\begin{acknowledgments}
 We are grateful to Professor M.~Ichimura for his 
helpful correspondence.
 We also acknowledge the dedicated efforts of
the RCNP cyclotron crew for providing a high
quality polarized proton beam.
 The experiment was performed at RCNP under 
Program Number E300.
 This research was supported in part by 
the Ministry of Education, Culture, Sports, 
Science, and Technology of Japan.
\end{acknowledgments}

\clearpage



%


\clearpage

\clearpage

\clearpage

\clearpage

\clearpage


%
\begin{table}
\caption{DWIA predictions of the polarization transfer
coefficients $D_{NN}(0^{\circ})$ and $D_{LL}(0^{\circ})$ 
for the ${}^{3}{\rm He}(p,n)3p$ reaction 
at $T_p$=346 MeV and $\theta_{\rm lab}$=$0^{\circ}$.\label{table:dwia}}
\begin{ruledtabular}
\begin{tabular}{crr}
$J^{\pi}$ & 
\multicolumn{1}{c}{$D_{NN}(0^{\circ})$} & 
\multicolumn{1}{c}{$D_{LL}(0^{\circ})$} \\
\hline
$1/2^+$ & $-0.17$ & $-0.56$ \\
$1/2^-$ & $-0.11$ &  $0.18$ \\
$3/2^+$ & $-0.15$ & $-0.65$ \\
$3/2^-$ &  $0.16$ & $-0.33$ \\
\end{tabular}
\end{ruledtabular}
\end{table}



\clearpage
\bibliography{e300}

\end{document}